# High Accuracy VLP based on Image Sensor using Error Calibration Method

Shihuan Chen[1],

Weipeng Guan[2]

1School of Automation Science and Engineering, South China University of Technology, Guangzhou 510640, China

2Department of Electronic and Computer Engineering, The Hong Kong University of Science and Technology, Hong Kong, China

* Co-Corresponding author: gwpscut@163.com

**Abstract** In this paper, visible light positioning (VLP) where the receiver adopts a commercial image sensor is considered. We firstly analyze the theoretical limits and error source of the VLP system using image sensor. And then, we develop a VLP positioning model on the receiver movement and further propose two novel error calibration algorithms, namely Rotation Calibration Method and dispersion circle calibration method. The rotation algorithm estimates the rotation center in the image instead of treating the image center as the rotation center, leading to reduced positioning error. For the dispersion circle, it can offset the shift error created by the conversation between different coordinate during the positioning calculation. According to the experimental results, the average positioning error of the proposed methods can be reduced to 0.82cm, which achieve state-of-the-art in the VLP field.

## 1 Introduction

There is a growing demand for real-time indoor navigation in many mobile applications, such as indoor service robots, indoor parking, indoor location-based services (LBS), etc. However, Global Positioning System (GPS) cannot be directly applied to indoor environment since the satellites' signal may be blocked by the exterior wall of buildings, while the traditional radio based indoor positioning technologies, such as Bluetooth, Wi-Fi, Radio-frequency Identification (RFID) and ultra-wideband (UWB), still have some disadvantages in terms of low accuracy, high latency, electromagnetic interference or high hardware cost.

Based on LED and visible light communication technologies, visible light positioning (VLP) has attracted more and more attention due to many advantages, including high positioning accuracy, low cost and dual functionality of illumination and positioning. Depending on the devices for receiving light signal, current VLP technologies can be divided into two categories: image sensor (IS) based VLP or photodiode (PD) [23、24] based VLP. PD is not an ideal positioning device for mobile terminals because the positioning based on PD will cause large errors due to the angle measurement, the received signal strength measurement, the light intensity variation and so on, and the accuracy of PD largely depends on the direction of beam. On the contrary, using image sensor as a receiver is a better choice. Firstly, visible light positioning based on image sensor does not have the above-mentioned defects of PD. Secondly, commercial mobile phones are usually equipped with image sensors, which makes the visible positioning system based on image sensors do not have to pay for the extra receiver and therefore further reduces the application cost of the visible positioning system.

Therefore, VLP technology, especially the VLP based on IS, has received significant interests from both academic and industrial fields. Figure 1 provides a comparison of some previous survey works [1~19] which have experimental verification. [26] have proposed several IS-based VLP systems using industrial camera, which can achieve high positioning accuracy of 1.417 with positioning latency lower than 80 ms. [27] can achieve an average positioning accuracy of 3.93 cm with moving speed up to about 38.5 km/h. [28] proposed the VLP algorithm based on machine learning to realize 3.65cm positioning accuracy. Although these researches have shown that accurate localization may be possible, little has been published about the theoretical limits. The determination of such theoretical limits will allow the optimization of the parameters governing IS-based VLP systems. And the error calibration method based on the analysis of theoretical limits, especially the one can be directly used in different VLP method, is also important for improvement of the positioning accuracy.

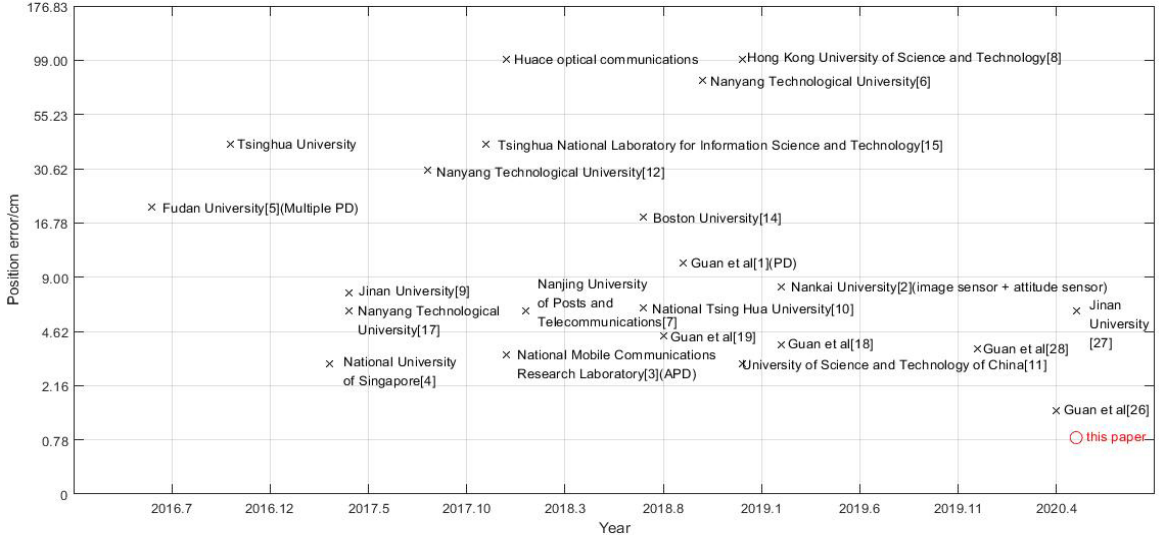

Fig. 1 Comparison of the positioning accuracy of VLP survey papers

In this paper, we firstly deeply analyzed the previous VLP works, dual-light positioning [18] and three light positioning [19], and figure out the theoretical limits of the VLP based on IS. After that, we novel propose two different optimization methods, Rotation Calibration Method and dispersion circle calibration method. More specifically, two design aspects are considered to reduce the positioning error. The first aspect is to estimate the rotation center in the image, which further reduces the positioning error compared with employing the image center as rotation center. The second one is the calibration of shift error through the proposed dispersion circle calibration method. The experiment shows that the positioning accuracy of the proposed optimized VLP system is greatly improved. As can be seen in figure 1, the experimental results in this paper are state-of-the-art in the same type of positioning algorithm. The rest of this article is organized as follows. The second part introduces the VLP system model and the proposed error calibration algorithm. The third part gives the experimental results and analysis of the system. And the final part is the conclusion.

## 2 Theory and Limits of VLP based on IS

### 2.1 Theoretical Model

In an VLP system based on IS, the positioning anchors are white LEDs, which are installed on the ceiling, provide illumination and broadcast their positioning signal simultaneously. Each VLP luminaire is assigned a unique identifier (UID) when it is manufactured. When it is installed, this UID will be associated with a unique set of coordinates, and the VLP system maintains a UID coordinates mapping table once all luminaires are in position. The receiver is a camera image sensor, such as that embedded in a robot moving around the room or in a smartphone held by someone. Both the signals from the LEDs and the images of the LEDs on the camera image sensor are used to determine the location and azimuth angle of the receiver.

The 3-D world coordinate system, 3-D camera coordinate system, and 2-D image plane coordinate system, are shown in figure 2. All of the LED $P_i = (X_i, Y_i, Z_i)^T (i = 1,2,...N)$ in the world coordinate, is mapped onto an image point $p_i = (x_i, y_i)^T (i = 1,2,...N)$ in the image coordinate through the lens of the camera. While the $p_i$ can be measured. Although the observation value of the imaging point is often influenced by noises [24], through the ROI detection method [26,29], it can be measured precisely and robustness

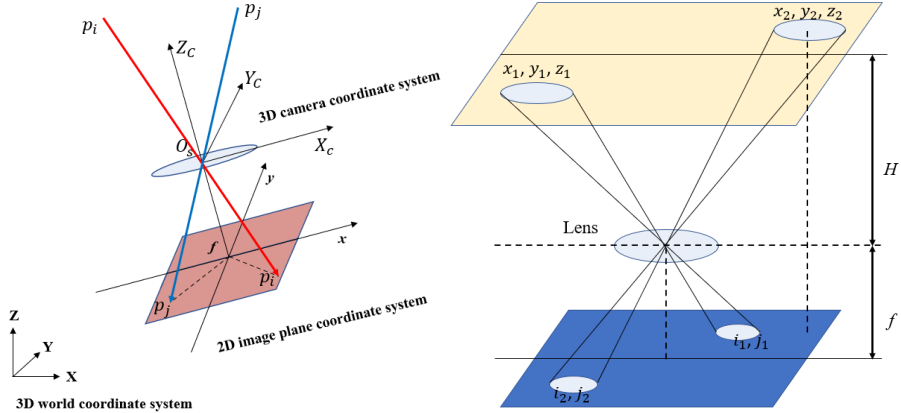

Fig. 2. VLP system model based on IS.

For any LED $P_i = (X_i, Y_i, Z_i{}^T (i = 1,2,...N$ in the world coordinate, we can model the VLP system model as:

$$P_{ci} = \boldsymbol{R}(P_i - O_s) \qquad (1)$$

Where $P_{ci} = (X_{ci}, Y_{ci}, Z_{ci}{}^T (i = 1,2,...N$ is the coordinate of LED $P_i$ in the 3-D camera coordinate system. $O_s = (X_s, Y_s, Z_s{}^T$ is the center of the camera, which also refer to the coordinate of the positioning terminals. $\boldsymbol{R}$ is the rotation matrix from the 3-D world coordinate system to the 3-D camera coordinate system, as following:

$$\boldsymbol{R} = R_x(\alpha) \bullet R_y(\beta) \bullet R_z(\gamma) = \begin{bmatrix} 1 & 0 & 0 \\ 0 & cos\alpha & sin\alpha \\ 0 & -sin\alpha & cos\alpha \end{bmatrix} \bullet$$

$$\begin{bmatrix} cos\beta & 0 & -sin\beta \\ 0 & 1 & 0 \\ sin\beta & 0 & cos\beta \end{bmatrix} \bullet \begin{bmatrix} cos\gamma & -sin\gamma & 0 \\ sin\gamma & cos\gamma & 0 \\ 0 & 0 & 1 \end{bmatrix} \qquad (2)$$

Where $\alpha, \beta$ and $\gamma$ is the angle along the X, Y, Z axis. The rotation angles $\alpha$ and $\beta$ can be directly read out from the inclination sensor attached to the receiver. If a robot is moving horizontally on the floor or a smartphone is placed face up, then only the azimuth angle $\gamma$ has to be considered, the other two rotation angles are taken to be equal to $0^o$. Then, the rotation matrix of the camera with respect to the 3-D world coordinate system can be expressed as

$$\boldsymbol{R} = R_z(\gamma) = \begin{bmatrix} cos\gamma & -sin\gamma & 0 \\ sin\gamma & cos\gamma & 0 \\ 0 & 0 & 1 \end{bmatrix} = \begin{bmatrix} a & -b & 0 \\ b & a & 0 \\ 0 & 0 & 1 \end{bmatrix} \qquad (3)$$

Where, $a = cos\gamma$, $b = sin\gamma$, and $a^2 + b^2 = 1$.

For the $Z_s$ of the positioning terminal, it can be calculated by:

$$Z_s = Z_i - H \qquad (4)$$

Where $H$ is the vertical distance from $O_s$ to the roof. Since the rotation angles of X and Y axis are taken to be equal to $0^o$, then vertical distance $H$ between the LED and the lens plane can be expressed as

$$\frac{H}{f} = \frac{D_{12}}{d_{12}} = \frac{\sqrt{(X_1-X_2)^2+(Y_1-Y_2)^2}}{P_d\sqrt{(x_1-x_2)^2+(y_1-y_2)^2}} \qquad (5)$$

Where $f$ is the focal length, $D_{12}$ is the physical distance between the two LEDs, $d_{12}$ is the distance between the LED pixels, and $P_d$ is the conversion of pixel distance and physical distance (which would be elaborated in section 2.2). $f$ and $P_d$ are the intrinsic parameter of the camera.

The relationship between 3-D world coordinate of LED $P_i$ in the 3-D camera coordinate system $P_{ci} = (X_{ci}, Y_{ci}, Z_{ci})^T$ and the LED in the image coordinate $p_i = (x_i, y_i)^T$ can be described by:

$$\frac{X_{ci}}{x_i} = \frac{Y_{ci}}{y_i} = \frac{Z_{ci}}{-f} \qquad (6)$$

Substituting (3), (4) and (6) into (1) we derive the mathematical relation between the LEDs, $(X_i, Y_i)_{i=1}^N$, and the observation values of their corresponding imaging points $(x_i, y_i)_{i=1}^N$, then we have

$$\begin{pmatrix} x_i \\ y_i \end{pmatrix} = \left(-\frac{f}{H}\right) \begin{pmatrix} a & b \\ -b & a \end{pmatrix} \begin{pmatrix} X_i - X_s \\ Y_i - Y_s \end{pmatrix} \qquad (7)$$

For the calculation of $R_z(\gamma)$, when we can obtain three LED in the view of the camera, we can obtain the simultaneous equations to calculate (the details can be seen in [19]). When only two LEDs are obtained in the view of the camera, we can calculate the angle between the vector formed by two LED points and the coordinate system (the details can be seen in [18]). Then we can realize the 3D positioning based on two or three LEDs using IS.

## 2.2 Theoretical Limits and Error Analysis

As what have been mentioned in section 2.1, we can conclude that there are four main reasons for the introduction of errors in VLP. The first and most overlooked one is the installation, which contribute to the error 3-D world coordinate of LED $P_i$. Secondly, the positions of the LEDs in the image plane $P_{ci}$ may suffer measurement error that will introduce additional error to the receiver positioning. What's more, there is an implicit process in the equation (1), which is the conversation between the camera coordinate and the image coordinate, it would have shift error. Last but not the least, as the fabrication error, which create rotation around the lens center, the center of the corresponding rotation on the image may not be exactly the center of the image plane. In other words, for the rotation around the lens center, the center of the corresponding rotation on the image plane may not be exactly the center of the center of the camera $O_s$.

As can be seen in figure 3, another coordinate which is called pixel coordinate is introduced based on the three coordinates (world coordinate system, camera coordinate system, image coordinate) in figure 2. The origin of image coordinate system is the point of intersection of camera's optical axis and the image sensor imaging plane, i.e., the center of midpoint of the image sensor imaging plane. The unit of the mentioned image coordinate system is mm, which belongs to physical unit. The unit of the pixel coordinate is pixel, which is described by its row and line. Firstly, after obtaining the pixel coordinates of the LEDs (through ROI detection algorithm), according to the relationship between the pixel coordinate system and the image coordinate system, the coordinates of the two LEDs in the image coordinate system can be calculated:

$$\begin{cases} x = (u - u_0)di \\ y = (v - v_0)dj \end{cases} \quad (8)$$

Where $x, y$ is the 2-D image plane coordinate system; $u, v$ is the 2-D pixel coordinate system. $di, dj$ represent unit transformation of two coordinate systems respectively, $1\text{pixel} = di\ mm$. $(u_0, v_0)$ is the midpoint of image coordinate in the pixel coordinate system. Therefore, we can obtain the physics distance between different LEDs on the image, and use the equation (5) for the positioning calculation. However, the origin of the image coordinate might not exactly at the center of the pixel coordinate. The main reason for this phenomenon lies in two aspects, the imperfect parallelism between the receiver and the LED planes, and the imperfect installation of the optical lens.

Generally, it is well known that the more LEDs are used as the anchor, the higher positioning accuracy it can be achieved. However, in our group previous experimental results [18,19] shows that the positioning accuracy of the double LEDs VLP algorithm (average error is 1.99 cm) is better than the three LEDs one (average error is 2.14 cm), in the similar experimental setup. According to the error analysis above, at first, the installation error of LEDs is eliminated, since both of them are the same setup and the installation error in such small environment can be ignored. What's more, through the ROI detection method [26,29], the observation value of the LED in image pixel coordinate can be measured precisely and robustness. Therefore, the main source of the error should be created by the conversation between the camera coordinate and the image coordinate, and rotation error. Both of calibration procedures will be elaborated in the remainder of this Section. As for the double LEDs VLP algorithm, only two LED are measured, which is one less than the three LEDs VLP algorithm. The less LED are measured, the less shift and rotation error would be introduced. This is explanation that why the experiment test doesn't match the common sense. Then we can also calibrate the error which is introduced into the VLP system.

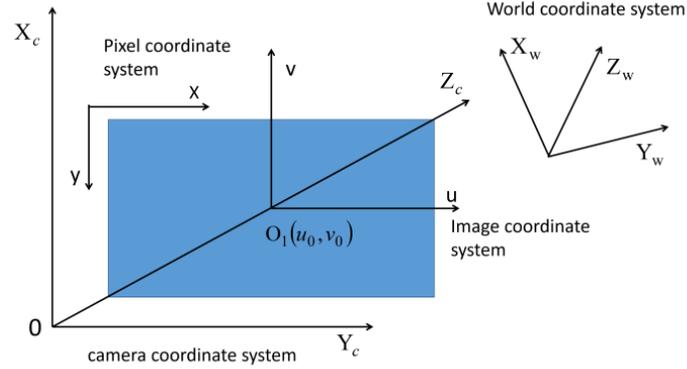

Fig.3. The xy is the pixel coordinate system, vu is the image coordinate system, Xc Yc is the camera coordinate system, and Xw Yw Zw is the world coordinate system.

## 3 Design of the Error Calibration Algorithms

### 3.1 Rotation Calibration Method

To avoid such positioning error, in this work we set a reference point, record the positions of the LEDs at the reference point, and estimate the receiver shift from the reference points according to the shift and rotation of the pixel regions of the LEDs. As shown in figure 4. $O_1(u_0, v_0)$ is the theoretical center of the pixel coordinate. To identify the rotation center, we select 12 angles among 0 to 360 degrees around the lens center, which are roughly in the order of each 30 degrees, but not strictly 30 degrees. Then, we can obtain $O_2(u_1, v_1)$ through calculating the circle center of the trajectory of the LEDs. It is observed that the center of the rotation (red point) is not strictly the image center as the black point.

The rotation center from the above calibration is adopted for the rotation step in the positioning process. Then, equation (8) can be modified into

$$\begin{cases} x = (u - u_1) \ di \\ y = (v - v_1) \ dj \end{cases} \quad (9)$$

However, even if we can obtain the rotation center to replace the image center for solving the rotation error, shift error still exists. Since the measure points of the rotation calibration process would not exactly match as a circle.

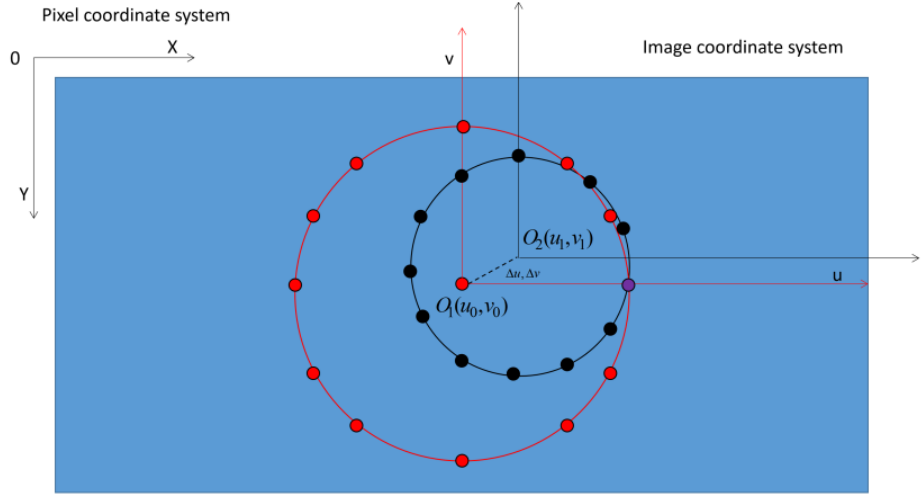

Fig. 4 Shift and rotation calibrated in the pixel and image coordinate

## 3.2 Dispersion Circle Calibration Method

Based on the problem mentioned in section 3.1, in this section, we proposed dispersion circle calibration method to deal with the shift error of the image center. We can assume that most of the positioning results of the system are in this circle. Ideally, when the center of the dispersion circle is corrected to coincide with the actual position coordinates, the Euclidean distance between the data and the center of the dispersion circle is the actual positioning error, and most of the error do not exceed the dispersion radius. As long as the dispersion radius is small enough, the results can be very accurate when the camera is precisely calibrated. As shown in figure 6, we place the image sensor at the origin point of the world coordinate system and make enough measurements to find the smallest circle that can include all the results. The position of the center of the circle is the center of the dispersion circle. In practice, because the positioning results are generally about the symmetrical distribution of the center of the dispersion circle, the average value of all the results is taken as the position of the center of the dispersion circle for timeliness. So we can get the deviation between the center of the dispersion circle and the actual position, which is shown in the following equation (10) (11).

$$\Delta x = x_c - 0 = \overline{x} = \frac{1}{n}\sum_{i=1}^{n} x_i \qquad (10)$$

$$\Delta y = y_c - 0 = \overline{y} = \frac{1}{n}\sum_{i=1}^{n} y_i \qquad (11)$$

Based on the deviation, we use the following formula to correct the pixel coordinates of the image sensor:

$$u_1 = u_0 + \frac{\Delta x}{di} \tag{12}$$

$$v_1 = v_0 + \frac{\Delta y}{dj} \tag{13}$$

Finally, we use the traditional method of camera calibration to establish the correspondence between the known coordinate points on the calibration object and its image points through the calibration object whose size is known so that we can obtain the internal and external parameters of the camera model.

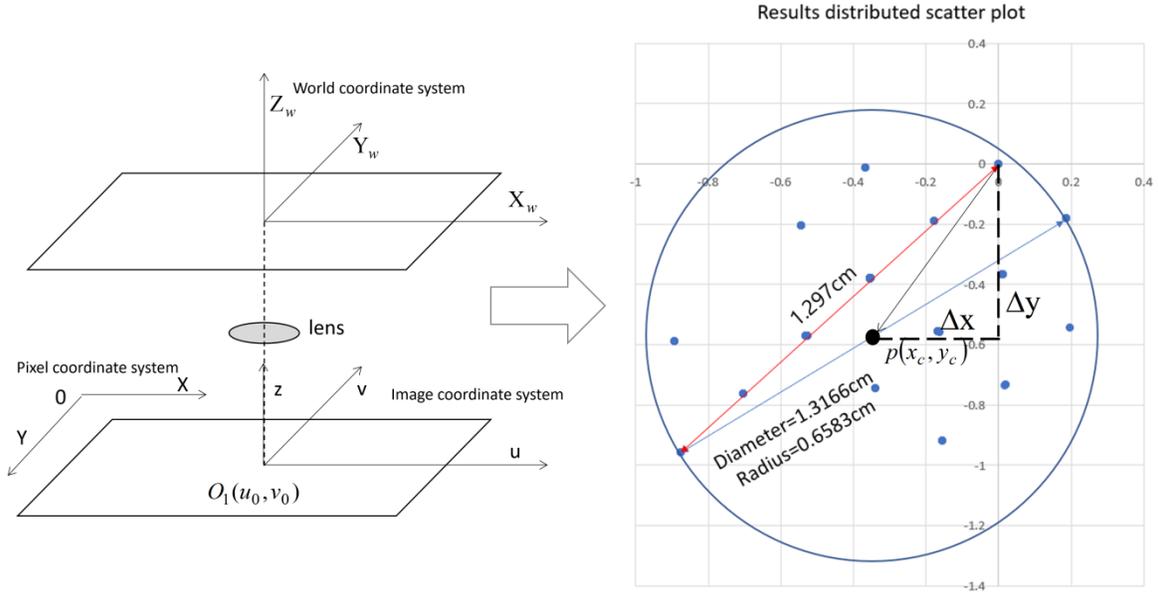

Fig. 6. Optimization schematic diagram based on dispersion circle.

## 4 Experiment

### 4.1 Experiment Setup

Experiments with field tests have been conducted to evaluate the positioning accuracy of the proposed indoor navigation system in terms of stationary positioning and dynamic positioning. The experimental platform and the related specifications are shown in figure 7 and Table 1. A remote control robot carrying a CMOS image sensor is used as the mobile unit in the experiments to test the real-time VLP performance in terms of positioning accuracy and positioning speed. The processor is Pi 3 Model B with Quad ARM Cortex-A53 Core 1.2 GHz Broadcom BCM2837 64 bit CPU and 1 GB RAM.

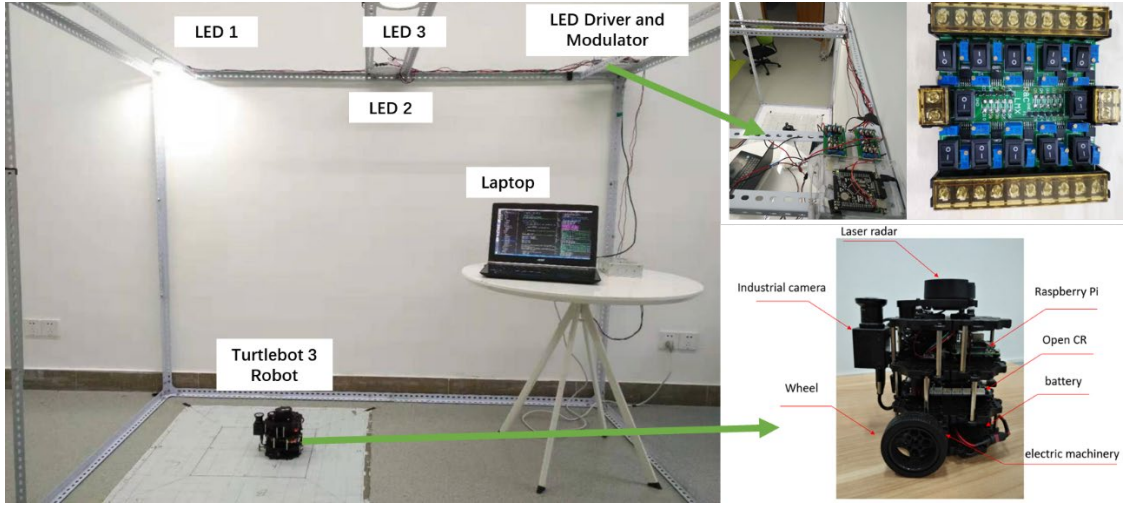

Fig.7 Experimental environment and hardware.

Table 1. Parameters in this paper.

| Parameter | Value |
|---|---|
| Indoor space unit size(L × W × H) (m 3 ) | 2 × 1.1 × 1.6 |
| The focal length (mm) | 3 |
| Height of the camera (m) | 0 to 0.3 (resolution: 0.1) |
| Plan range of the camera (m) | 0.1 to 0.7 (resolution: 0.2) |
| Voltage of each LED (V) | 28.43 |
| Current of each LED (A) | 0.1 |
| The resolution of the camera | 800 × 600 |
| The exposure time of the camera (ms) | 0.05 |
| Computer parameter | Acer Aspire VN7-593G,Intel (R) Core (TM) i7-7700HQCPU@ 2.8GHz, Ubuntu 16.04 LTS |

### 4.2 Positioning Accuracy

To evaluate the positioning accuracy of the proposed VLP system, two series of experiments were carried out.

#### 4.2.1 Stationary Positioning Accuracy

The first series were used to test the performance for motionless objects. First, we measure the actual position of the camera point by means of plumb and coordinate paper. Then, the image processing program segmented the LED image taken by the camera to obtain the position of the LED in the pixel coordinate system and used the linear classifier to identify LED-ID [20~22] to obtain the position of the LED in the world coordinate system. Finally, the positioning results are obtained through the positioning algorithm. Without loss of generality, both of the proposed error calibration: rotation calibration method and dispersion circle calibration method, are integrated into the previous works: the double-LED VLP algorithm [18] and three-LED VLP algorithm [19]. As shown in Fig. 8 and Fig. 9, 36 spots with a grid pattern inside the experimental area were selected and the standard deviation between the measured positions and the actual positions were calculated. Each position is tested 12 times, then totally 432 positioning results are obtained.

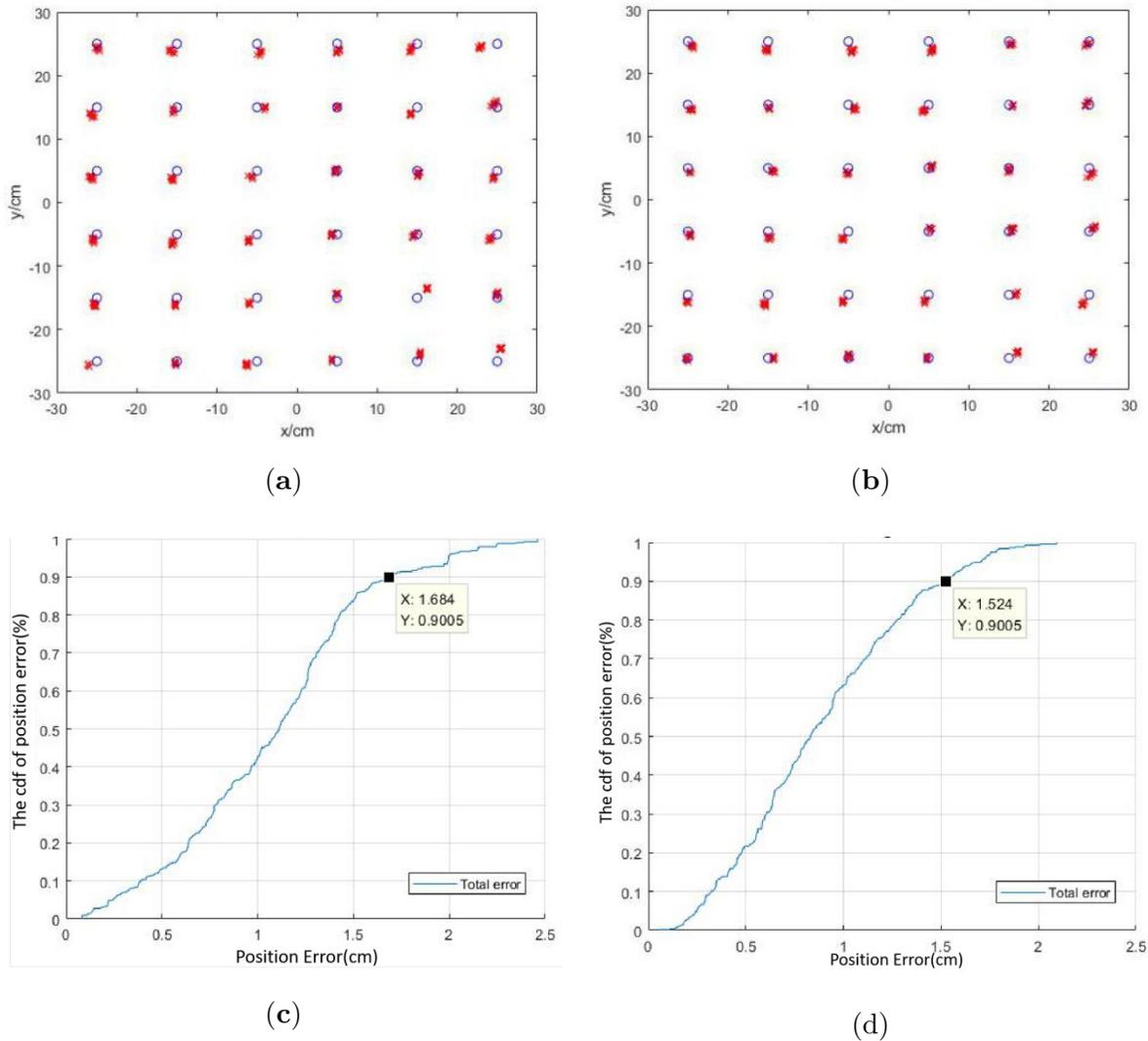

Fig. 8 The positioning result of VLP based on three LEDs. (a) using the proposed rotation calibration method; (b) using the proposed dispersion circle calibration method; (c) and (d) The CDF curves of the rotation calibration method and dispersion circle calibration method, respectively;

As can be seen in figure 8, the average positioning error of the three-LED VLP using the proposed rotation calibration method is 1.08cm, while the average positioning error of the three-LED VLP using the proposed dispersion circle calibration method is 0.88 cm, compared with the average error of 2.14 cm in [19] in larger experimental setup.

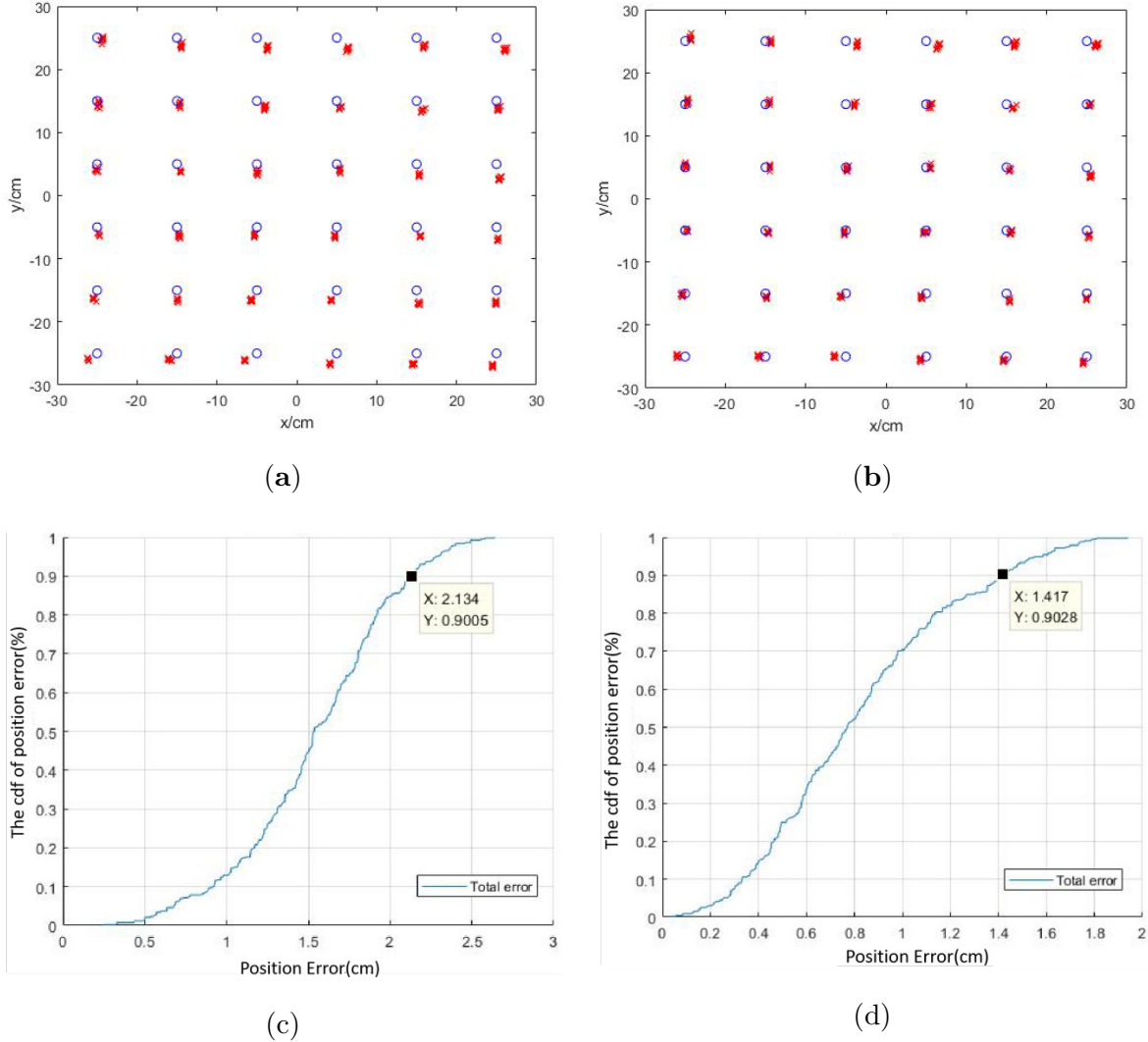

Fig. 9 The positioning result of VLP based on two LEDs. (a) using the proposed rotation calibration method; (b) using the proposed dispersion circle calibration method; (c) and (d) The CDF curves of the rotation calibration method and dispersion circle calibration method, respectively;

As can be seen in figure 8, the average positioning error of the double-LED VLP using the proposed rotation calibration method is 1.54cm, while the average positioning error of the double-LED VLP using the proposed dispersion circle calibration method is 0.82 cm, compared with the average error of 1.99 cm in [18] in larger experimental setup. Through comparing the double-LED VLP (more than 90% test point, less than 2.13cm) and three-LED VLP ((more than 90% test point, less than 1.68cm)) using the proposed rotation calibration method, we can get a conclusion that, as for the proposed rotation calibration method, the more LEDs can be obtained, the higher accuracy can be achieved, since more LED information are introduced as the calibration. However, as for the proposed dispersion circle calibration method, not only it can achieve similar accuracy in different systems, but also achieve better performance than the proposed rotation calibration method.

### 4.2.2 Dynamic Positioning Accuracy

The second series of experiments were used to test the performance for moving mobile units. In this section, we just test the double-LED VLP using the proposed dispersion circle calibration method to verify the dynamic performance. As for the mobile robot which is moving, it is really hard to know the exact actual position of the robot without using complexity motion capture system. As we just give a command for straight movement at a fixed speed to the robot in our experiments, which means that if the initial position and orientation of the robot are not exactly aligned, the actual trajectory of the robot would deviate from the given command trajectory. Then the estimated trajectory of the VLP algorithm using the error calibration method would deviate from the given command trajectory, which is a normal experimental phenomenon.

Therefore, we record the starting position and the ending position of the robot to represent the actual trajectory of the robot. We calculate the positioning error through comparing the command trajectory, actual trajectory of the robot and the estimated trajectory estimated by the proposed VLP algorithm using the error calibration method, which can be seen in figure 10 (a) and (b). The robot is moving along the given command trajectory (the green line) at a speed of 4 cm/s, while blue line is the actual trajectory of the robot obtained by measuring the start and end points of the robot. The yellow line is the estimated trajectory of the proposed VLP algorithm using the error calibration method, which is obtained by fitting the VLP estimated results with the least square method. To evaluate the accuracy of the dynamic positioning, we sampled the test points from the green line and the yellow line. More specifically, as the exact actual position of a moving robot is not available, we just only use the command trajectory which controls the actual trajectory of the robot as the reference to calculate the dynamic positioning errors. Although the actual trajectory of the robot has errors comparing the given command, the errors between the VLP estimated trajectory and the command trajectory is still persuasive. The CDF of the dynamic positioning errors can be seen in figure 10 (c) and (d). It is worth to mention that, the position results of the VLP algorithm using the error calibration method is the coordinates of the center of CMOS camera, which can be converted to the center of the robot through the Transform Frame (TF) transformation.

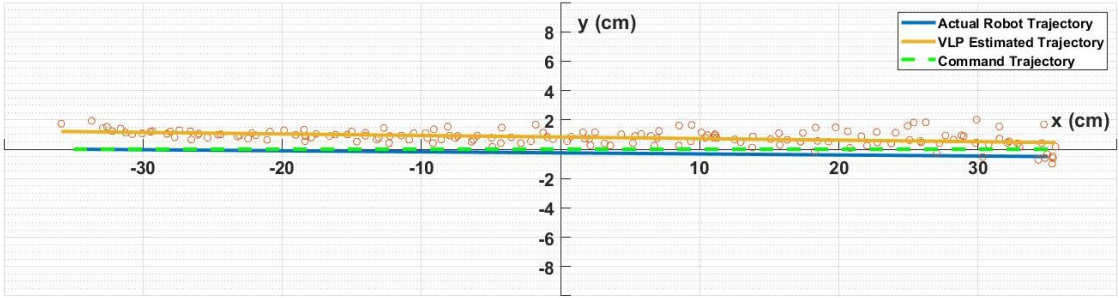

(a)

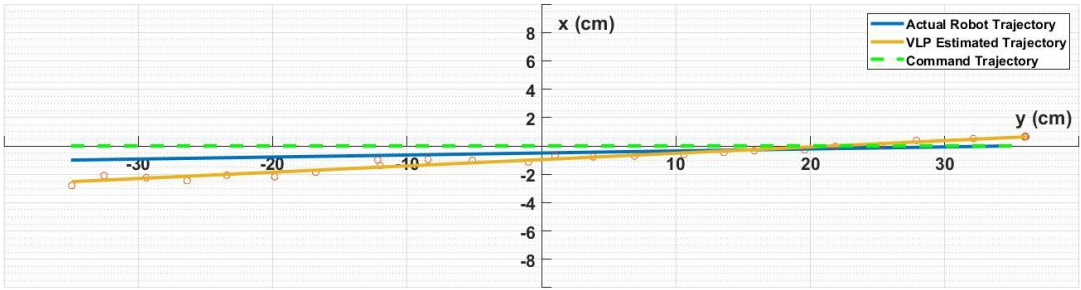

(b)

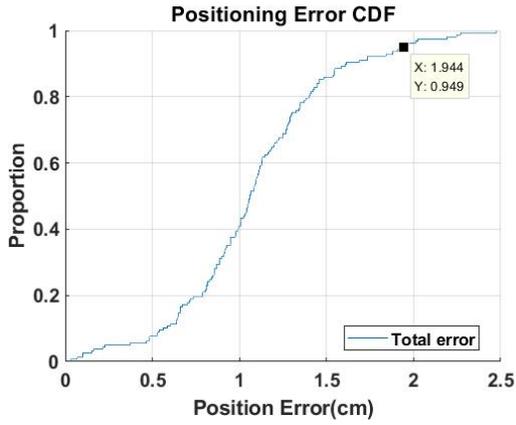

(c)

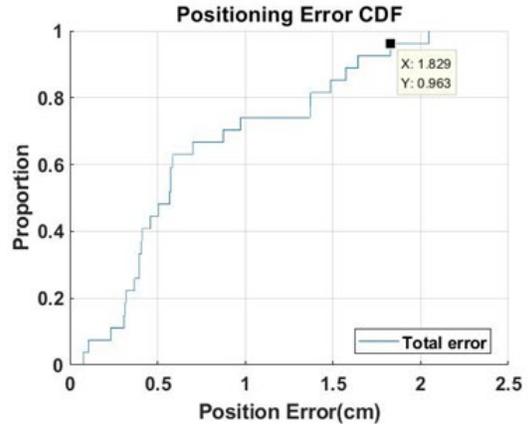

(d)

Fig.10. The accuracy of dynamic positioning of the double-LED VLP using the proposed dispersion circle calibration method. (a) The experiment result of robot moving from (-35, 0) to (35, 0); (b) The experiment result of robot moving from (0, -35) to (0, 35); (c) and (d) The CDF curves of positioning errors in (a) and (b), respectively.

# 5  Conclusion

In this paper, we have proposed a centimeter-level accuracy positioning system based on a smartphone camera and visible light LEDs. Two error calibration algorithms, namely Rotation Calibration Method and dispersion circle calibration method are proposed. Experiments are carried out to test the performance for motionless objects and moving mobile units. Experimental results show that the proposed system can provide the average positioning error of 0.82cm with the maximum positioning error of 1.93cm, which achieve state-of-the-art in the VLP field.

# Acknowledgments

This work is an upload version of the following papers:

https://arxiv.org/ftp/arxiv/papers/1911/1911.11773.pdf

# References

[1] Jiang J, Guan W, Chen Z, et al. Indoor high-precision three-dimensional positioning algorithm based on visible light communication and fingerprinting using K-means and random forest[J]. Optical Engineering, 2019, 58(1): 016102.

[2] Ji Y Q, Xiao C X, Gao J, et al. A single LED lamp positioning system based on CMOS camera and visible light communication[J]. Optics Communications, 2019, 443: 48-54.

[3] Guo Z, Jia Z, Xia W, et al. Indoor localization system for mobile target tracking based on visible light communication[C]//2017 IEEE 85th Vehicular Technology Conference (VTC Spring). IEEE, 2017: 1-5.

[4] Zheng H, Xu Z, Yu C, et al. A 3-D high accuracy positioning system based on visible light communication with novel positioning algorithm[J]. Optics Communications, 2017, 396: 160-168.

[5] Xu Y, Zhao J, Shi J, et al. Reversed three-dimensional visible light indoor positioning utilizing annular receivers with multi-photodiodes[J]. Sensors, 2016, 16(8): 1254.

[6] Zhang R, Zhong W D, Qian K, et al. A reversed visible light multitarget localization system via sparse matrix reconstruction[J]. IEEE Internet of Things Journal, 2018, 5(5): 4223-4230.

[7] Zhu B, Cheng J, Yan J, et al. VLC positioning using cameras with unknown tilting angles[C]//GLOBECOM 2017-2017 IEEE Global Communications Conference. IEEE, 2017: 1-6.

[8] Zhou B, Liu A, Lau V. Robust Visible Light-Based Positioning Under Unknown User Device Orientation Angle[C]//2018 12th International Conference on Signal Processing and Communication Systems (ICSPCS). IEEE, 2018: 1-5.

[9] Fang J, Yang Z, Long S, et al. High-speed indoor navigation system based on visible light and mobile phone[J]. IEEE Photonics Journal, 2017, 9(2): 1-11.

[10] Li Y, Ghassemlooy Z, Tang X, et al. A VLC smartphone camera based indoor positioning system[J]. IEEE Photonics Technology Letters, 2018, 30(13): 1171-1174.

[11] Xu J, Gong C, Xu Z. Experimental indoor visible light positioning systems with centimeter accuracy based on a commercial smartphone camera[J]. IEEE Photonics Journal, 2018, 10(6): 1-17.

[12] Zhang R, Zhong W D, Kemao Q, et al. A single LED positioning system based on circle projection[J]. IEEE Photonics Journal, 2017, 9(4): 1-9.

[13] Zhang S, Zhong W D, Du P, et al. Experimental demonstration of indoor sub-decimeter accuracy VLP system using differential PDOA[J]. IEEE Photonics Technology Letters, 2018, 30(19): 1703-1706.

[14] Little T, Rahaim M, Abdalla I, et al. A multi-cell lighting testbed for VLC and VLP[C]//2018 Global LIFI Congress (GLC). IEEE, 2018: 1-6.

[15] Zhao C, Zhang H, Song J. Fingerprint and visible light communication based indoor positioning method[C]//2017 9th International Conference on Advanced Infocomm Technology (ICAIT). IEEE, 2017: 204-209.

[16] Park J K, Woo T G, Kim M, et al. Hadamard matrix design for a low-cost indoor positioning system in visible light communication[J]. IEEE Photonics Journal, 2017, 9(2): 1-10.

[17] Zhang R, Zhong W D, Qian K, et al. Image sensor based visible light positioning system with improved positioning algorithm[J]. IEEE Access, 2017, 5: 6087-6094.

[18] Guan W, Zhang X, Wu Y, et al. High Precision Indoor Visible Light Positioning Algorithm Based on Double LEDs Using CMOS Image Sensor[J]. Applied Sciences, 2019, 9(6): 1238.

[19] Guan W, Wen S, Liu L, et al. High-precision indoor positioning algorithm based on visible light communication using complementary metal–oxide–semiconductor image sensor[J]. Optical Engineering, 2019, 58(2): 024101.

[20] Guan W, Li J, Wen S, et al. The Detection and Recognition of RGB-LED-ID Based on Visible Light Communication using Convolutional Neural Network[J]. Applied Sciences, 2019, 9(7): 1400.

[21] Li J, Guan W. The Optical Barcode Detection and Recognition Method Based on Visible Light Communication Using Machine Learning[J]. Applied Sciences, 2018, 8(12): 2425.

[22] Xie C, Guan W, Wu Y, et al. The LED-ID detection and recognition method based on visible light positioning using proximity method[J]. IEEE Photonics Journal, 2018, 10(2): 1-16.

[23] Guan W, Wu Y, Wen S, et al. Indoor positioning technology of visible light communication based on CDMA modulation[J]. Acta Opt. Sin, 2016, 36(11): 66-74.

[24] Zhao X, Lin J. Theoretical limits analysis of indoor positioning system using visible light and image sensor[J]. Etri Journal, 2016, 38(3): 560-567.

[25] Guan W, Wen S, Zhang H, et al. A Novel Three-dimensional Indoor Localization Algorithm Based on Visual Visible Light Communication Using Single LED[C]//2018 IEEE International Conference on Automation, Electronics and Electrical Engineering (AUTEEE). IEEE, 2018: 202-208.

[26] Guan W, Chen S, Wen S, et al. High-Accuracy Robot Indoor Localization Scheme based on Robot Operating System using Visible Light Positioning[J]. IEEE Photonics Journal, 2020, 12(2): 1-16.

[27] Lin P, Hu X, Ruan Y, et al. Real-time visible light positioning supporting fast moving speed[J]. Optics Express, 2020, 28(10): 14503-14510.

[28] Chen Y, Guan W, Li J, et al. Indoor Real-Time 3-D Visible Light Positioning System Using Fingerprinting and Extreme Learning Machine[J]. IEEE Access, 2019, 8: 13875-13886.

[29] Guan W, Liu Z, Wen S, et al. Visible Light Dynamic Positioning Method Using Improved Camshift-Kalman Algorithm[J]. IEEE Photonics Journal, 2019, 11(6): 1-22.